# Durable and Recoverable Hydrophilicity of Polyethylene Terephthalate Fabric Prepared with Plasma Selective Etching

Tarig T. A. Owad [a, d], Eshraga A. A. Siddig [e], Renad E. M.Salih [a, d], Yu Zhang [a], Yu Xu [a,b,c,*], Jing Zhang [a,b,c,*]

[a] New Energy Materials and Devices, Applied Physics Department, College of Science, Donghua University, Shanghai 201620, People's Republic of China

[b] Textile Key Laboratory for Advanced Plasma Technology and Application, China National Textile & Apparel Council, Shanghai 201620, People's Republic of China

[c] Member of Magnetic Confinement Fusion Research Center, Ministry of Education, Shanghai 201620, People's Republic of China

[d] College of Science, Sudan University of Science and Technology (SUST), Khartoum, 11111, Sudan

[e] College of Engineering, Sudan University of Science and Technology (SUST), Khartoum 11111, Sudan

*Corresponding author: J. Zhang (E-mail: jingzh@dhu.edu.cn), Yu Xu (E-mail: yuxu@dhu.edu.cn)

## ABSTRACT

Durable delustered PET (PET-TiO$_2$) fabrics super hydrophilic surface has been obtained by plasma selecting etching. The aging effect of their hydrophilicity after plasma treatment has been investigated with storage time. After Ar/O$_2$ radio frequency (RF) plasma treatment for only 7 min, PET-TiO$_2$ fabric showed water contact angle of 0º. After 10 month storage time, it keeps its water contact angle below 75.7º. Further more, with Xenon light irradiation for 10 min, it is firstly found that it has well-recovered water contact angle to 5°. While the contact angle of PET fabric for 7 min returns to 123.0° and its hydrophilicity disappeared almost completely and showed no response to Xenon light irradiation. The water absorption rate of 7 min plasma treated PET-TiO$_2$ fabric increased by 57.54%. By field emission scanning electron microscopy (FE-SEM), X-ray photoelectron spectroscopy (XPS) and X-ray diffraction analysis(XRD) measurement, waviness structure of humps and ridges with irregular particles or pits were found on the plasma treated PET-TiO$_2$ fabric surface and increased Ti atomic percentage was observed. It is verified that TiO$_2$ particles inside PET-TiO$_2$ fiber have been exposed to its surface by plasma selective etching of its organic component. It suppresses the aging effect and is characterized with durable and recoverable hydrophilicity. This one step, quick, green and cost-resonable manufacture method has a pratical application for durable superhydrophilic surfaces.

## 1. Introduction

As an important kind of commodity, polyethylene terephthalate (PET) fabric is commonly used in the textile industry as clothing, sports, packaging, smart, electronic, furniture, automotive materials. PET fabric has high performance and many physical, mechanical, and chemical advantages, such as durability, excellent washing performance, light-weight, high tensile strength, flexibility, high transparency, quick-drying, reduced-wrinkles and shrinkage rate, excellent thermal and chemical stability, good dimensional stability, resistance to many environmental conditions and UV radiation as well as low cost [1-6]. However, its inert molecular structure and low moisture resorption rate make it have a poor hydrophilicity and uncomfortable for wearing [7, 8]. It has long been explored to induce chemical and physical changes to PET fabric surfaces to improve its wettability without affecting its bulk properties [9-

11].

Several chemical techniques have been applied to fabricate hydrophilic PET fabrics and change its hand feeling, texture drape, dyeing depth and fastness such as alkali treatment [12], acid treatment [11], aminolysis and graft polymerization etc.. However, except the degradation of its bulk properties and the impact on its mechanical properties, the high energy consumption, environmental pollution and long period of time are the main concerns to its sustainable applications [13-15].

Plasma treatment is a widely studied method for modifying the surface characteristics of organic materials with little damage to its bulk properties [16-18]. PET fabric can be made hydrophilic through plasma treatments by introducing highly reactive species on its surface. It contains a lot of reactive species such as positive and negative ions, UV photons, electrons, excited molecules and atoms, free radicals, which can break down covalent bonds in polymers and start many surface chemical reactions [19]. The properties of the treated surfaces are determined by the precursor gases from non-polymeric (He, Ar, $H_2O$, $O_2$ etc.) or polymeric gases( $CH_2CHCOOH$, HCHO etc.) and reaction parameters (frequency, power, time, pressure, flow rate etc.) [20, 21]. The emergence of hydroxyl (—OH), carboxyl (—COOH), and carbonyl groups (=C=O) are primarily promoted by $O_2$ plasma. Besides, plasma can be employed for nano-scale texture of fiber surface in dry state. Interestingly, plasma etching on fabric surfaces will etch away the organic component and make rough surface through breaking out its molcular bonds by ionization or dissociation [12, 22]. The introduction of polar functional groups or short graft chain and rough surface make it easy absorption of water or other vapor solvent immediately after plasma treatment. Yu Ren et al [23] used a dielectric barrier discharge (DBD) plasma treatment on the nanostructure, wettability, and hydrophilic recovery of polylactic acid (PLA) surfaces. Elabid et al [24] used atmospheric pressure Ar/$O_2$ plasma to tailor the surface morphology, chemical structure and enhance the low-temperature dyeability of disperse dyes on PET fabric. Zhang et al [25] reported the atmospheric-pressure air/He plasma to enhance PET absorbance for water-based pigment inks. Weclawski et al [26] investigated the effects of combining atmospheric plasma with UV excimer laser exposure of nylon 66 fibre surfaces to a novel diagnostic dyeing process in the presence or absence of additional UV excimer laser exposure. Syromomtina et al. [27] used a low-temperature RF plasma (oxygen and ammonia) to modify poly(3-hydroxybutyrate) P3HB thin films and improve their hydrophilic properties by forming polar functional group.

Despite these beneficial features, still there exist some disadvantages of plasma treatments for preparing super-hydrophilic surfaces on polymers. After storage for some time in air after plasma treatment, the polymers become hydrophobic again because the polar functional groups saturate with the environmental atmosphere or reverse inside for surface energy balance. Therefore, the hydrophilicity or activity of treated polymer surfaces may wane with storage time and so-called aging effect of plasma treatment has been problem over time [28, 29]. PET fabrics and foils were modified by low-pressure RF plasmas with air, $CO_2$, water vapor as well as Ar/$O_2$ and He/$O_2$ mixtures and the best hydrophilization effect with respect to water suction and aging is obtained by Ar/$O_2$ plasma treatment [30]. The ageing process is related with the storage conditions and can be suppressed by storing the PET films at low temperature or by decreasing the relative humidity of the surrounding air [31]. The wettability of high density polyethylene (HDPE) powders within less than 0.1 s Ar/$O_2$ plasma surface modification is correlated with the [O]/[C] atomic ratio and the polar surface tension and the ageing proceeds mostly within the first 7 days of storage [32].

Inorganic semicondutive nanoparticles (such as $TiO_2$, ZnO or $SiO_2$ nanoparticles) is usually mixed into melted PET polymer matrix and spinned into PET-TiO$_2$, PET-SiO$_2$ or PET-ZnO hybride polymer fiber to give it unique functional properties. TiO$_2$ is a semiconductor with a bandgap of 3.2 eV and high refractive index. It is a powerful hydrophilic additive and white pigment which is environment friendly, low cost, thermally stable, source-rich and anti-abrasion. Compared with the vapor coating or sol-gel coating on the fabric surface, the mixing of the inorganic TiO$_2$ nanoparticle with the polymer matrix is easy and their combing is strong enough and refer the polymer with special UV absorption, silky appearance, soft hand feeling, texture drape properties or wettability [33]. According to the amount of TiO$_2$ nanoparticles mixed into PET polymer, 'clear' or 'semi-dull' or 'dull' (delustered) PET fibers are obtained popularly for different uses [34]. However, the TiO$_2$ nanoparticles inside the PET matrix can not display the same functionality as in its powder form, especially the role of its surface-related functionality. Because TiO$_2$ particles have been wrapped around with PET polymers and their surface are covered with a very thin polymer layer, the surface-related or contacting-related properties of TiO$_2$ can not be exhibited well enough.

In this paper, we developed a one-step and simple technique to prepare a durable and recoverable PET fabric surface using delustered PET-TiO$_2$ sample with plasma selective etching. Both pure PET fabric and delustered PET fabric were employed and treated by plasma. However, the hydrophilicity of the plasma-treated pure PET disappeared almost completely after 10 month storage. While plasma treated delustered PET fabric showed much better hydophilicity retention. TiO$_2$ nanoparticles in delustered PET fabric have been verified to be exposed to its fiber surface by plasma etching. Additionally, with Xenon light irradiation for several minutes, the super-hydrophilicity of the treated delustered PET fabric is recovered at once and its water contact angle was kept between $5^0$ to $13.8^0$ during further monthly periodic irradation.

## 2. Experimental

### 2.1. Materials

The plain weave pure PET (weight ∼50 g m$^{-2}$) and delustered PET fabrics (weight ∼64.7 g m$^{-2}$) with about <2.0% weight percentage TiO$_2$ nanoparticles were provided by Hua Fang Co. Ltd., China. The fabric samples were washed in distilled water in an ultrasound bath for 30 min at 80°C, then dried in a 70°C oven for 1 hour. The cleaned fabrics were cut into 25 cm × 25 cm dimensions for the experiment. Argon and oxygen (Shanghai Chenggong Gas Industry Co. Ltd., 99.99%) were used as discharge gases.

### 2.2. Plasma treatment process

A capacitor (AP-600/300TM Nordson, March, USA) coupled radio frequency plasma with 13.56 MHz power source was used for the surface etching and operated in vacuum [35]. The samples were placed between the upper and lower parallel shelf electrodes with a 5 cm gap. Oxygen and Argon were supplied into the reactor and the working pressure was kept at 280 mTorr. The mass flowmeters were used to control the flow rate of O$_2$ (93 sccm) and argon (7 sccm). The plasma treatment was conducted at 375W power with different etching times (0.5 ~ 7 min). As we all know, TiO$_2$ nanoparticles is a photo-active material and demonstrated a strong super-hydrophilic behavior under irradiation of light [36]. So, after storage in air at room temperature, the samples were irradiated with a 300W Xenon lamp (Perfect light. China) with 1 m distance. The samples are named with different experimental conditions and the * means samples with 10 min light irradiation shown in Table 1.

**Table 1.** The sample names and conditions used in the experiment

| Treatment Method | Plasma etching | | | | Xenon light irradiation | |
|---|---|---|---|---|---|---|
| | 0 min | 1 min | 3 min | 7 min | 10 min | 10 min |
| Samples Name | PET | PET-1 | PET-3 | PET-7 | PET* | PET*-7 |
| | PET-TiO$_2$ | PET-TiO$_2$-1 | PET-TiO$_2$-3 | PET-TiO$_2$-7 | PET-TiO$_2$* | PET-TiO$_2$*-7 |

2.3. Surface characterization

Field emission scanning electron microscopy (FE-SEM, S-4800, Hitachi Ltd., Japan) was adopted to observe the surface morphology of the samples. The chemical states of sample surfaces were identified through X-ray photoelectron spectroscopy (XPS) with a Kratos Axis Ultra DLD XPS (Thermo Fisher ESCALAB 250Xi, USA) using an Al K, X-rays radiation source (1253.6 eV). For XPS spectra processing, Avantage software was employed to deconvolve and analyze the C1s, O1s, and Ti2p peaks. For crystalline structure analysis, the X-ray diffraction (XRD) experiments were completed on a Rigaku diffractometer (D/max-2550 PC, Japan) with a revolving anode and a Cu k$_\alpha$ radiation source ($\lambda=1.54056\times10^{-10}$ m). For the test, a current of 150 mA and a voltage of 40 kV were used. A Dropmeter$^{TM}$ professional A-200 device (Ningbo Haishu Meishi Detection Technology Company, China) was used to measure the water contact angles with 5 μL of deionized water.

**3. Results and Discussion**

3.1 Morphology

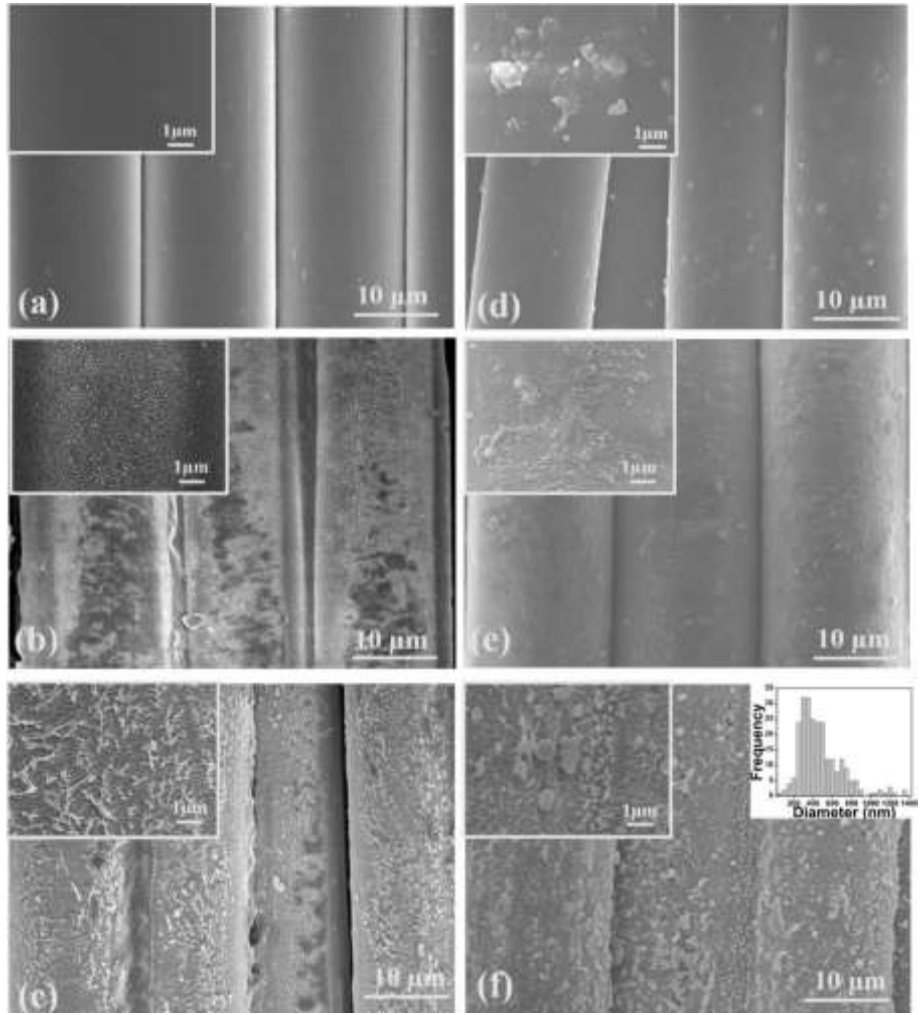

**Figure 1.** The FE-SEM images of (a) PET, (b) PET-1, (c) PET-7, (d) PET-TiO$_2$, (e) PET-TiO$_2$-1, (f) PET-TiO$_2$-7 and the inset is the particles distribution.

Fig. 1 shows the morphology changes of the samples by using FE-SEM. Fig. 1(a) shows that the surface of the untreated PET fabric sample without TiO$_2$ nanoparticles is smooth. Following the plasma treatments for 1 min, the PET fabric surface presents a slight roughness on the surface, as shown in Fig. 1(b). The roughness increases with the time increasing to 7 min and the surface is structured with aligned forming rows of humps and ridges in Fig. 1(c). The etched structure were created when Ar/O$_2$ plasma active species as (Ar$^*$, Ar$^+$, O$^-$, O$_2^+$, O$^+$ etc.) reach and react with the polymer surface [37]. The organic component is etched away through breaking down the polymer covalent bonds and active groups like —COOH, —C=O, —C—OH were introduced onto the fiber surface [17, 38]. The waveness structure is also due to the partly cyrstallined domains of PET fibers uniaxially drawn with high speed and the different etching ratio of crystalline to amorphous phase [39].

The surface of the untreated PET-TiO$_2$ fabric was rough with some sparely distributed nanoparticles as in Fig. 1(d). After 1 min plasma etching treatment, the PET-TiO$_2$-1 fabric sample surface became rougher with small etching nanopits and nanoparticles as shown in Fig. 1(e). It is estimated that the nanopits and nanoparticles are exposed TiO$_2$ from the inner of the fiber to its outer surface as we will verify later. As shown in Fig. 1(f), the roughness of the PET-TiO$_2$-7 fabric surface increased and showed similiar waviness structure composed of humps and ridges as in pure PET-7 but with more identified pits. The size of the main particles was calculated and the distribution was shown in the inset. These irregular pits and particles were visible with a larger size than those observed in Fig. 1(e), which could be attributed to TiO$_2$ nanoparticles protruding out of the surface.

As investigated in our previous studies through atomic force microscope (AFM) of pure PET and PET-TiO$_2$ fabric surface before and after Ar/O$_2$ plasma treatment, the root mean square roughness (RMS) of pure PET and PET-TiO$_2$ increased from 4.2 nm to 23.71 nm and 22.35 nm to 45.61 nm successively [40]. The obvious variation of RMS of treated PET and PET-TiO$_2$ indicates that their etching process is different. Because of the etching rate differences of organic PET polymer and inorganic TiO$_2$ in Ar/O$_2$ atmosphere, the surface organic PET thin layer covered on TiO$_2$ particles are etched away and TiO$_2$ nanoparticles are exposed to the surface. Then the exposed TiO$_2$ nanoparticles works as a mask for prohibiting further etching of the polymer matrix. This selective etching of organic PET polymers creates a higher RMS surface with large amount of exposing TiO$_2$ nanoparticles.

## 3.2. Chemical composition

The chemical composition of the samples were investigated by XPS. The high-resolution spectra of C1s, O1s, and Ti2p of PET and PET-TiO$_2$ with plasma treatment were shown in Fig. 2 and Fig. 3. The atomic concentration of C1s, O1s, and Ti2p are listed in Table 2.

C1s peaks of PET and PET-TiO$_2$ were deconvoluted to three peaks of carbon-containing groups and appeared at 284.76 (C–C/C–H), 286.40 (C–O/COH), and 288.8 eV (O–C=O/COOH), and shown in Fig. 2(a)-(c) and Fig. 3(a)-(c), respectively. With the treatment time increasing, the intensity of C–C/C–H decreases, while C–O and C=O increase. This corresponds to the introduction of more polar oxygen groups like C–O, COH, O-C=O and COOH on the fabric surface after plasma treatment.

Fig. 2(d)-(f) show that the spectra of O1s for PET were divided into two peaks, 531.77 eV and 533.31 eV. As in Fig 3 (d)-(e), the O1s peaks of PET-TiO$_2$ and PET-TiO$_2$-1 samples appeared at 531.68 eV and 533.25 eV. Morever, a new O1s peak of PET-TiO$_2$-7 sample appeares at about 530.05 eV. The peaks at 531.77 eV and 531.68 eV represented O-C=O, and the peaks at 533.31 eV and 533.25 eV were attributed to C-O. Following plasma treatment of 1 min and 7min, the atomic ratio of [O]/[C] for PET increased from 46.42% to 50.22% and increased from 52.48% to 56.60% for PET-TiO$_2$, respectively as shown in Table 2. The increased O1s intensity of O-C=O and C–O peaks is well corresponds to increased C1s band intensity of O-C=O /COOH, C–O / C=O above [24]. The O1s new peak at 530.05 eV of PET-TiO$_2$-7 sample indicates an increase in the surface is belong to Ti-O.

The XPS spectra of Ti2p peaks are shown in Fig 2 and Fig 3. The pure PET samples are Ti2p free even after different plasma treatment. Mealwhile, the PET-TiO$_2$ fabric shows obvious Ti2p peaks after 7 min plasma treatment. It contains three Ti2p deconvolution peaks: the peak at 458.9 eV attributed to $Ti^{4+}\ 2p^{3/2}$, the peak at 464.7 eV ascribed to $Ti^{4+}\ 2p^{1/2}$ and the peak at 457.9 eV due to the $Ti^{3+}\ 2p^{1/2}$ caused by the oxygen vacancies in the surface layer by the plasma exposure [41]. Since the organic component of the PET-TiO2 fabric shares the different etch rate to the $TiO_2$ particles, it are easy removed and the particles remain on the fibers surface. The XPS results above are summarized in Table 2.

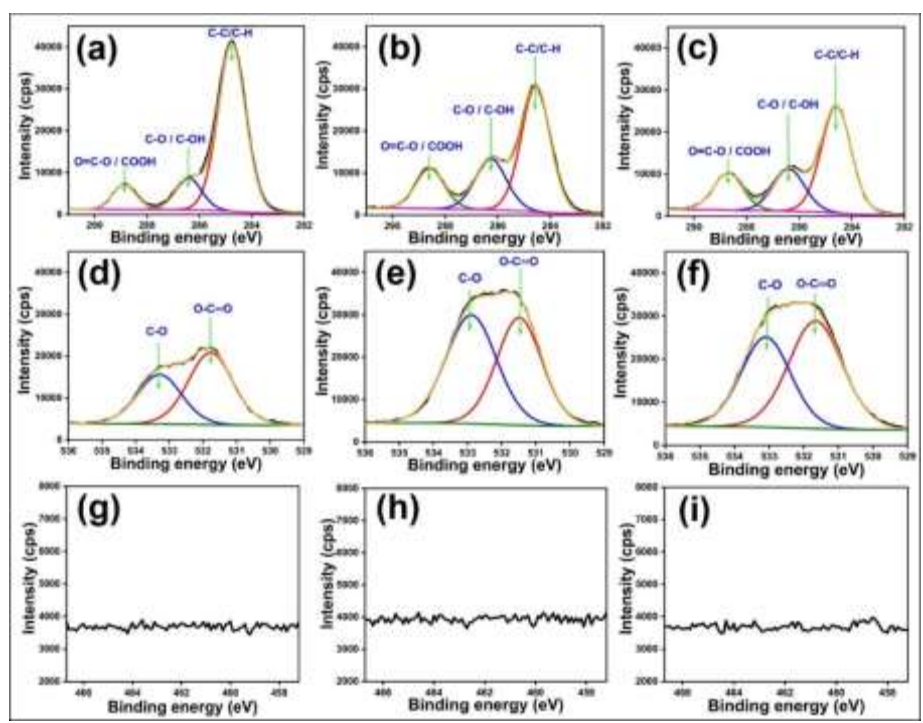

**Figure 2.** Deconvoluted XPS spectra of untreated and plasma-treated PET-1 and PET-7 fabrics: (a), (b), and (c) C1s; (d), (e), and (f) O1s; (g), (h), and (i) Ti2p; respectively.

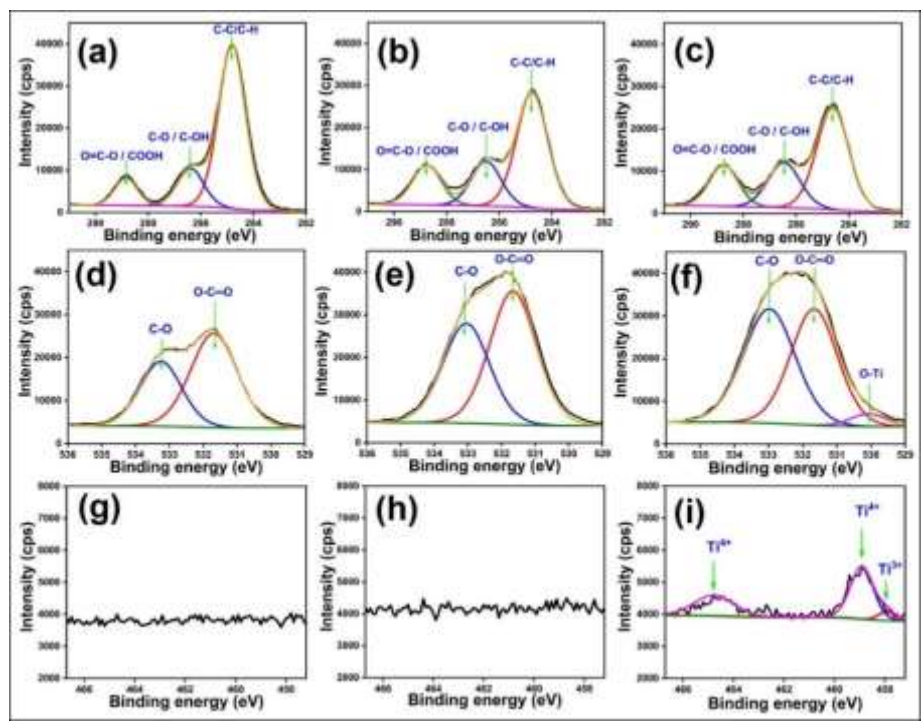

**Figure 3.** Deconvolution XPS spectra of untreated and plasma treated PET-TiO$_2$-1 and PET-TiO$_2$-7 fabrics: (a), (b), and (c) C1s; (d), (e), and (f) O1s; (g), (h), and (i) Ti2p.

**Table. 2.** Atomic concentration of samples.

| Sample | Atomic concentration (%) | | | Atomic ratio | |
|---|---|---|---|---|---|
| | C1s | O1s | Ti2p | [O]/[C] | [Ti]/[C] |
| PET | 78.52 | 19.88 | - | 25.31 | - |
| PET-1 | 67.04 | 31.16 | - | 46.47 | - |
| PET-7 | 65.24 | 32.8 | - | 50.27 | - |
| PET- TiO$_2$ | 73.91 | 24.47 | - | 33.10 | - |
| PET- TiO$_2$-1 | 64.06 | 33.62 | - | 52.48 | - |
| PET- TiO$_2$ -7 | 62.47 | 35.36 | 0.70 | 56.60 | 1.12 |

The XRD patterns of the samples were shown in Fig 4. In the PET sambles XRD patterns, TiO$_2$ peaks are missed even after plasma treatment. Meanwhile, TiO$_2$ in PET-TiO$_2$ and PET-TiO$_2$-7 fabric were detected. Peaks were found at 25.4°, 38.1°, and 48.2°, which relate to crystalline phase (101), (004), and (200) of anatase TiO$_2$, respectively, verifying the crystal phase with a standard XRD pattern of anatase TiO$_2$ (JCPDS files No. 21-1272) [42].

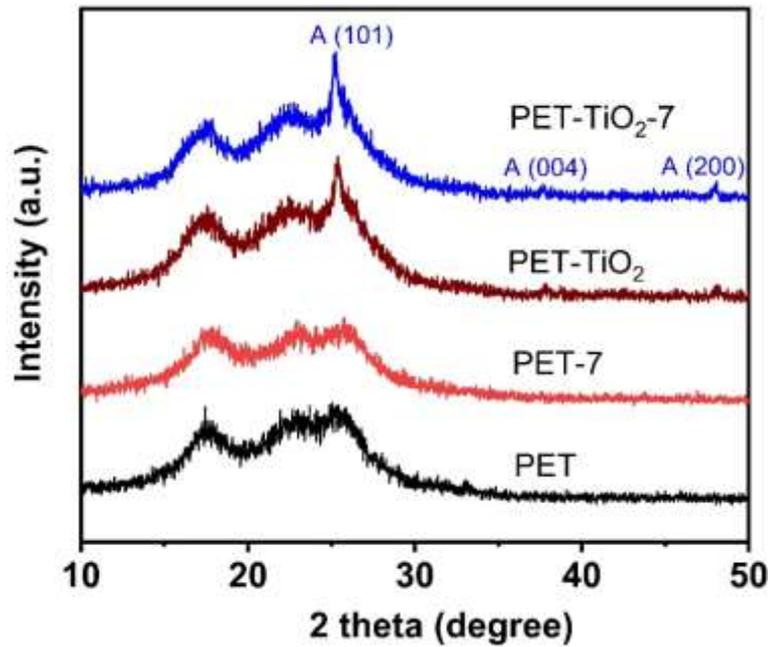

**Figure 4.** XRD pattern of the pure PET and PET-TiO$_2$ samples before and after plasma etching.

3.3. Wettability of PET and PET-TiO$_2$ fabrics with different treating methods.

As shown in the XPS results, plasma etching introduce hydrophilic groups on the surface, which may caused it turns to hydrophilic. The static water contact angle measurement before and after plasma treatment immediately are shown in Fig. 5. The original PET and PET-TiO$_2$ fabrics are 133.6° and 122.8°, respectively. The plasma-treated pure PET fabric samples for 0.5 min, 1 min, 3min and 7 min decreased to 44.3°, 23.5°, 12.3 ° and 0°, respectively. Meanwhile, PET-TiO$_2$ fabric with the same treated time are found to be 19.1°, 0°, 0°, 0°, respectively. This means that plasma treatment for improving hydrophilicity is more effective for PET-TiO$_2$ fabric than for pure PET fabric.

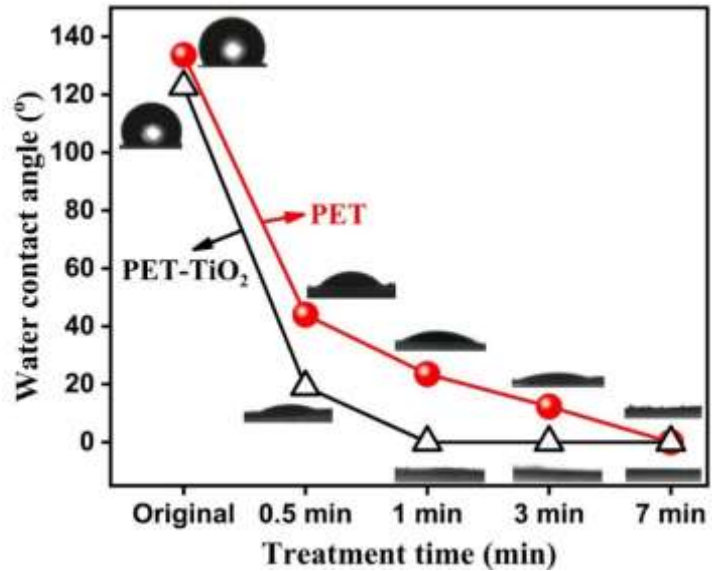

**Figure 5.** Water contact angles of PET and PET-TiO$_2$ fabric samples with different plasma treatment time.

In this study, the plasma etched fabric samples of PET-7 and PET-TiO$_2$-7 displayed the similar aging effect are shown in Fig. 6 by static water contact angle measurement. Immediately following plasma etching, the contact angles of both PET-7 and PET-TiO$_2$-7 fabrics considerably decreased to 0°. Storage in room temperature atmosphere to 10 months, a gradual increased trend of hydrophobic nature was observed for PET-7 and PET-TiO$_2$-7 fabric samples with time. However, the PET-TiO$_2$-7 fabric sample shows smaller rise rate and more stable hydrophilicity than PET-7 fabric sample. After 10 months, the contact angle of PET-7 fabric returns to 123.0°, while that of the PET-TiO$_2$-7 fabric returns to 75.7°. This means a durable and different aging process of PET-TiO$_2$-7 sample. After plasma treatment, PET fabric undergoes non-permanent physical and chemical changes on its surface [43]. Plasma-etched surfaces have a tendency to reverse their changes and return to their original chemical state over time [44] in a phenomenon known as ageing effect and become hydrophobic again. Surface functional groups activated by plasma have a short shelf life because they reconfigure to lower energy states (not durable/stable) [21, 45]. Increasing the treatment time and power may result in a more intensive creation of these groups within the surface layer, as well as a deeper surface modification of the sample surface. Other treatment or storage conditions have also shown influences on the aging effects, which may slow but never stops the aging process [46, 47]. It is also interesting to find that the wettability of the PET-TiO$_2$-7 fabric stored for 10 months was recovered after 10 min Xenon light irradiation, as shown in Fig. 6. The water contact angle of PET-TiO$_2$*-7 fabric decreases from 75.7° to about 5° and becomes superhydrophilic. This displays that the plasma aging effect has been suppressed to a great extent by combining with short time Xenon light irradiation on PET-TiO$_2$-7 fabric. However, the water contact angle of PET*-7 fabric decreased from 123° to about 119.9° only, which is not significantly changed and still shows hydrophobic character.

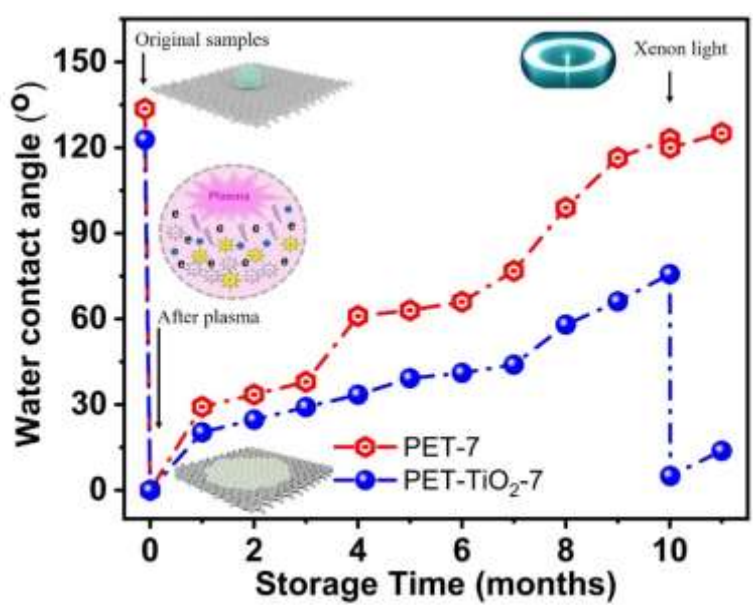

**Figure 6.** Water contact angle variation of untreated and plasma-treated PET and PET-TiO$_2$ fabric with storage time and Xenon light irradiation.

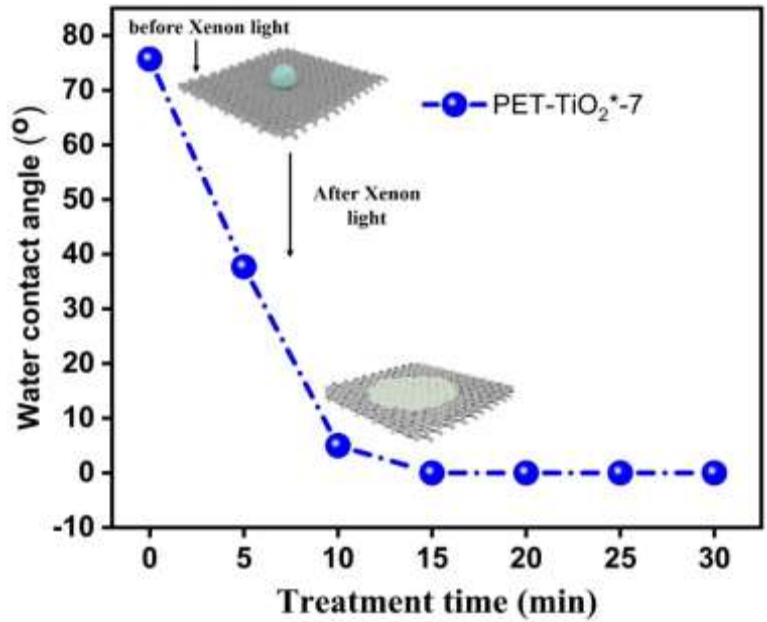

**Figure 7.** Water contact angle variation of PET-TiO$_2$*-7 fabric with different Xenon light irradiation.

Water contact angle variation of PET-TiO$_2$*-7 fabric with different Xenon light irradiation were investigated and shown in Fig. 7. After 5 min irradiating, PET-TiO$_2$*-7 fabric decreases from 75.7° to about 37.7°. Interestingly, after 10 min under irradiation of light, it shows a super-hydrophilic surface, and its water contact angle values were decreased from 37.7° to about 5°. The water contact angle is not further decresasd after the irradiation time increases to 15, 20, 25 and 30 min.

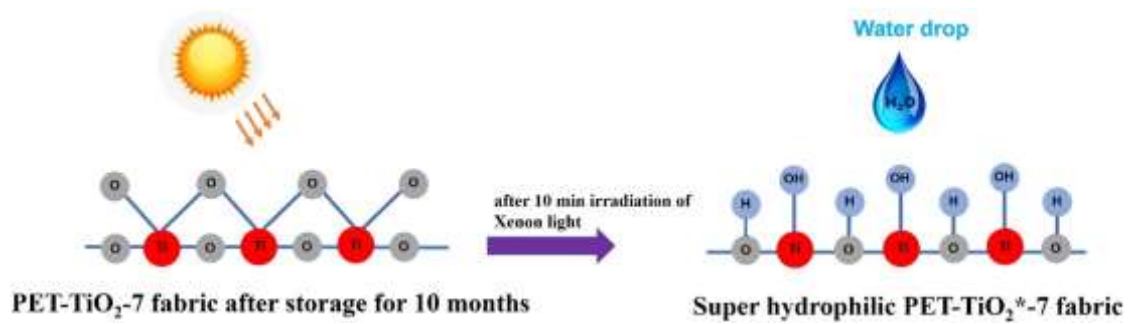

**Figure 8.** Superhydrophilicity mechanism on the surface exposed by TiO$_2$ particles.

TiO$_2$ is famous of light-induced hydrophilicity [48]. Electron and hole pairs are produced on the TiO$_2$ surface due to exposed TiO$_2$ nanoparticles activated by light irradiation, which will react with H$_2$O molecules via a strong electrostatic complex.

Ultraviolet irradiation may create surface oxygen vacancies at bridging sites, resulting in the conversion of relevant $Ti^{4+}$ sites to $Ti^{3+}$ sites [49] which are favourable for dissociative water adsorption [50, 51]. Accordingly, PET-TiO$_2$*-7 fabric surface displayed superhydrophilicity, as shown in Figure 8.

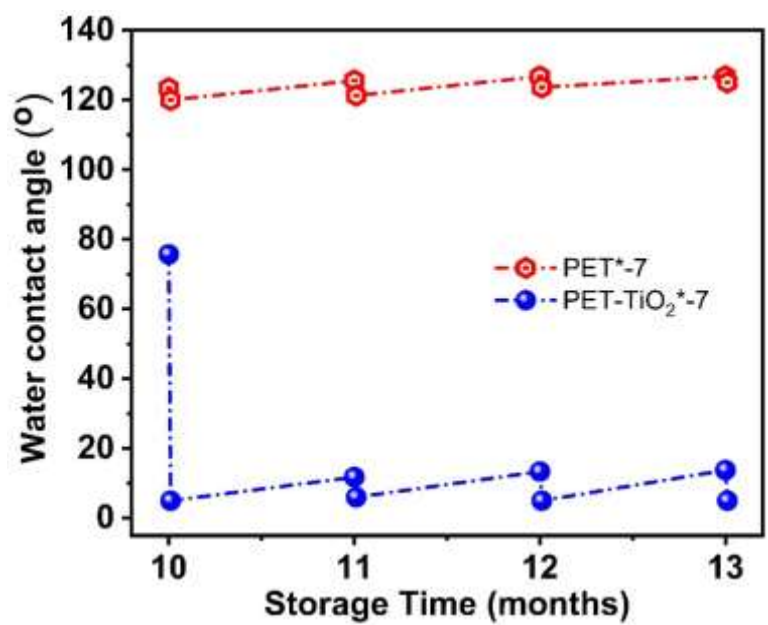

**Figure 9.** Water contact angle variation of PET*-7 and PET-TiO$_2$*-7 fabric with storage time during irradiation cycles.

Further monthly periodic irradiation of Xenon light on pure PET-7 and PET-TiO$_2$-7 fabric storaged for 10 months was done to compare the light effects on their recovery super-hydrophilicity. The water contact angles of the samples were tested every month before and after irradiation of 10 min. Water contact angles during irradiation cycles were shown in Fig. 9. The water contact angle of PET*-7 fabric increased slowly during the one month further storage time but it has almost no reply after irradiation. The water angles of PET-TiO$_2$*-7 fabric decreases to 5° and increases to about 13.8° after 1 month storage. The sample show well recoverable light-induced hydrophilicity in the following cycle test.

XPS of PET*-7 and PET-TiO$_2$*-7 storaged for 10 months were measured before and after the first 10 min Xenon light irradiation cycle for samples. No obvious moving of the peak position of C1s, O1s, and Ti2p and their XPS curves are omitted here. The atomic concentrations of C1s, O1s, and Ti2p are listed in Table 3. It is noticed that there is a very small changes of C1s, O1s, and Ti2p atomic concentration before and after 10 min Xenon light irradiation. However, two important comparision results sould be indicated. One is that the changes of O1s and T2p atomic concentration before and after 10 min Xenon light irradiation is much smaller than that undergoing 7 min plasma etching both for PET and PET-TiO$_2$ samples. These results display that plasma etching is more effective for the structure and composition changes for PET-TiO$_2$ than Xenon light in our experiment. The other one is that the atomic concentration of O1s for PET-7 and PET-TiO$_2$-7 all decreased than those 10 month ago. But the O1s atomic concentration of PET-7 decreased much more and is still lower than that for PET-TiO$_2$-7 after 10 month storage. The exposed TiO$_2$ nanoparticles on PET-TiO$_2$-7 fabric surface make its O1s keep at a higher atomic concentration and response to the Xenon light irradiation effectively [52, 53].

Therefore, in the case of PET-TiO$_2$ fabric, its wetting behavior and aging effect is mainly influenced by the exposed TiO$_2$ nanoparticles, introduction of polar oxygen groups and increased roughnes. As we know, TiO$_2$ is a large bandgap photocatalyst. It exhibits superhydrophilic surface behavior, together with a water contact angle value of zero under UV light [36, 54]. In theory, TiO$_2$ can be activated and generate electron-hole pairs. Under the full spectrum of 220-1200 nm of Xenon light, the photo-generated electrons can convert the Ti$^{4+}$ state to the Ti$^{3+}$ state and the holes react at bridging sites, resulting in oxygen vacancies [55]. Water molecules in the air occupy the oxygen vacancies, producing OH radicals and making the surface more hydrophilic. Because

Ti$^{3+}$ introduction and its narrowed bandgap by plasma etching, PET-TiO$_2$-7 fabric exhibites improved durable and recoverable super-hydrophilicity under irradiation of Xenon light.

**Table. 3.** Surface atomic concentration of samples with plasma and light.

| Sample | Atomic concentration (%) | | | Atomic ratio | |
|---|---|---|---|---|---|
| | C1s | O1s | Ti2p | [O]/[C] | [Ti]/[C] |
| PET-7 | 65.24 | 32.8 | - | 50.27 | - |
| PET-7 (10 months later) | 69.65 | 27.71 | - | 39.78 | - |
| PET*-7 | 71.1 | 26.04 | - | 36.62 | - |
| PET-TiO$_2$-7 | 62.47 | 35.36 | 0.70 | 56.60 | 1.12 |
| PET-TiO$_2$-7 (10 months later) | 65.97 | 32.19 | 0.66 | 48.79 | 1.00 |
| PET-TiO$_2$*-7 | 65.55 | 32.70 | 0.71 | 49.88 | 1.08 |

As etched PET-TiO$_2$ fabrics also exihibite improved water absorption rate :

$$WA\% = \frac{Ww - Dw}{Dw} \times 100\%$$

Where WA is the water absorption rate%, Ww is the dipped weight in g, and Dw is the dry weight in g. In Figure 10a, it is shown that WA% increases with the plasm etching time and changes from 70.5% to 81.16% for pure PET and 77.92% to 93.03% for PET-TiO$_2$ at the processing time of 1min and 7min, respectively. In order to illustrate how water absorption rate is increased after plasma etching, improvement rate (I%) was calculated using the water absorption rate% in accordance with the following equation:

$$I\% = \frac{W1 - W0}{W0} \times 100\%$$

Where $W_1$ and $W_0$ are, respectively, the weight of the treated and untreated fabric samples. In Figure 10b, it is observed that the improvement rate is 46.1% and 57.54% respectively for pure PET-7 and PET-TiO$_2$-7, comparing with its untreated corresponding fabric samples. Except to the polar oxygen functional groups, this result is also attributed to the increased RMS and large amount of exposed TiO$_2$ nanoparticles after plasma etching.

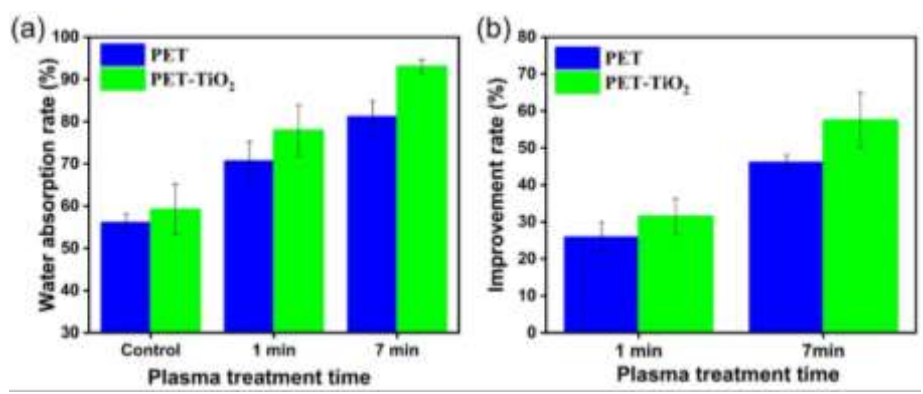

**Figure 10.** The water absorption rate (a), and improvement rate (b) of samples, before and after plasma treatment.

3.4. Suggested mechanism of the durable and recoverable superhydrophilicity of the surface

Figure 11 shows a flow diagram of the durable and recoverable superhydrophilic PET-TiO$_2$*-7 fabric preparation process. The entire process can be divided in two steps: low-pressure Ar/O$_2$ RF plasma surface etching treatment and Xenon light irradiation.

When PET-TiO$_2$ fiber surface is being treated with O$_2$/Ar plasma, a lot of physical and chemical reactions are expected to occur on the surface by the combardment of ions, high-energy electron, photons, radicals and other metastable species. These active species can lead to the chain scission or dissociation of the polymer covered on the top of TiO$_2$ particles or around them, which will etch and extract away the organic components especially their amorphous phases, which creates surface etching patterns like in Fig. 1(e),(f). Because of higher etching rate of Ar/O$_2$ plasma on organic PET than on inorganic TiO$_2$, further etching proceeds selectively around TiO$_2$ nanoparticles and makes TiO$_2$ be exposed on the surface with higher RMS. Ar/O$_2$ plasma also introduces reactive groups and produces large amounts of highly reactive polymer chain ends, such as free radicals, excited molecular oxygen and some oxidized polar anionic groups on the fiber surface like in Fig. 2 and Fig. 3. Although the etching rate on $TiO_2$ is slow, the RF plasma etching also has the tendency to generate oxygen vacancies and $Ti^{3+}$ on $TiO_2$ nanoparticles as displayed in Fig. 3 for further exposed treatment time. Because of the rougher surface, exposed $TiO_2$ nanoparticles, polar oxygen groups and $Ti^{3+}$ caused by plasma etching, PET-$TiO_2$ fabric displayed superhydrophilicity after 7 min plasma treatment. PET-$TiO_2$-7 fabric kept higher O1s atomic concentration and showed improved and durable hydrophilicity under the room atmosphere until 10 months. However, the hydrophilicity of treated pure PET is almost lost at all after 10 month storage time because of the active polymer chain reorientation inward and its O1s atomic concentration decrease.

In the second step of Xenon light irradiation, the XPS results show that Xenon light irradiation does not change the compositions of PET fabric obviously. But the PET-$TiO_2$-7 fabric surface with exposed anatase $TiO_2$ nanoparticles and $Ti^{3+}$ defect state is able to make use of the full spectrum of the Xenon light. Once PET-$TiO_2$-7 fabric surface has been excited by Xenon light irradiation, electron and hole pairs are formed on the PET-$TiO_2$*-7 fabric, which induces photocatalytic activity of $TiO_2$ nanoparticles. $TiO_2$ nanoparticles will react with $H_2O$ molecules. Hydrophilicity is enabled by activated sites with a polarity that prefer to attract $H_2O$ molecules via a strong electrostatic complex. Based on this, the dense clusters of hydroxyl groups can keep their surface charges and attract $H_2O$ molecules within the complex when the number of Ti-OH groups is sufficient. When the samples contain a large amount of donor-acceptor complex, the most upper surfaces become polar, resulting in long-term hydrophilicity [56-58]. Therefore, the exposed $TiO_2$ by Ar/$O_2$ plasma eching is most important for the durable superhydrophilicity under room atmosphere storage and its recovery under Xenon light irradiation.

The aging effect of PET-TiO$_2$-7 and PET-TiO$_2$*-7 is caused not only by the further polymer chain orientation and the Ti-OH groups relax, but also by the organic contamination absorbed on TiO$_2$ nanoparticle surface. However, the exposed TiO$_2$ can recover surface charge polarity and regain its hydrophilicity under Xenon light irradiation [59]. Therefore, a durable and recoverable superhydrophilicity surface is prepared.

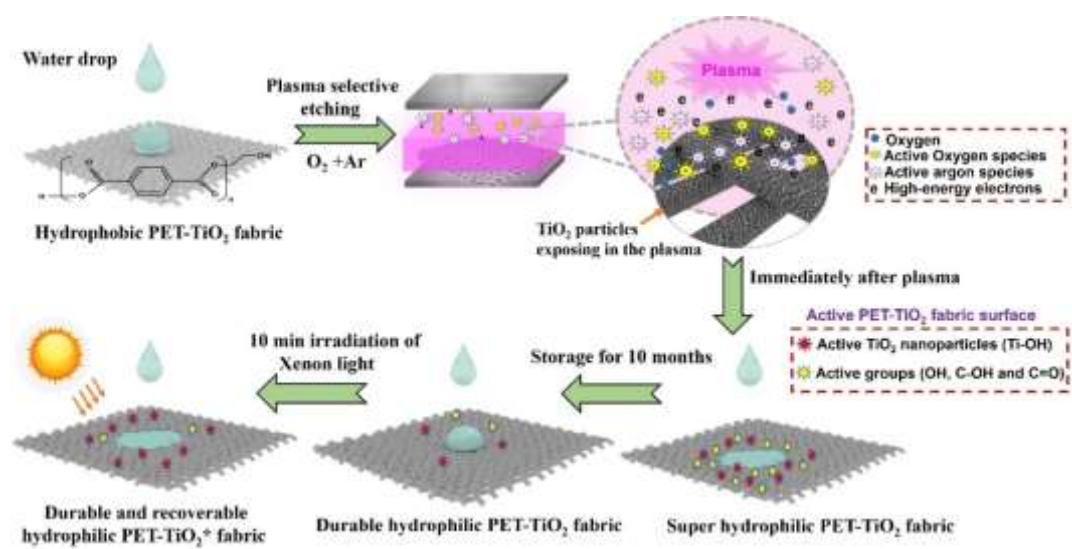

**Figure 11.** Schematic diagrams for the mechanism of preparing and recoverable superhydrophilic surfaces of delustered PET with plasma selective etching and Xenon light irradiation.

## 4 Conclusion

In summary, we demonstrated an efficient, simple, and environmentally friendly method for producing durable and recoverable superhydrophilic delustered PET fabric surface by plasma selected etching. First, by using delustered PET fabric and across plasma etching treatment, part of organic components have been selectively removed. The FE-SEM revealed a micro/nano waviness structure of humps and ridges on the fabric surface, along with exposed outstanding TiO$_2$ nanoparticles. XPS results show an increase of O1s atomic concentration and introduction of polar oxygen groups. The delustered PET fabric samples became more hydrophilic than PET fabric after the same time plasma etching treatment. After 10 months of storage, it still maintained the water contact angle at about 75°, showing better anti-aging performance. Interestingly, after Xenon light irradiation for only 10 min, the storaged PET-TiO$_2$ fabric for 10 months shows a well-recovered superhydrophilic surface immediately, and its water contact angle was kept at about 5°~13.8° during the monthly irradiation cycles. It is verified that the TiO$_2$ nanoparticles exposed on PET fiber surface by plasma selective etching are the main causes of the durable and recoverable superhydrophilic delustered PET fabric. This study provides an environmental friendly and excellent potential benign technique which can construct a durable hydrophilicity of PET fabric surface and could be easy recovered by light even after 10 month storge.

**Acknowledgements:**The National Natural Science Foundation of China, (No. 12075054, 11875104) and the Fundamental Research Funds of Central Universities (No. 2232021G-10, 2232019A3-12).